\documentclass[12pt,preprint]{aastex}

\def\lya{Lyman-$\alpha$}  

\def\erg{\hbox{erg}}

\def\ha0{$6559$\AA}
\def\h16{$6730$\AA}
\def\f17{f_{17}}

\def\kms{\hbox{km s$^{-1}$}}
\def\year{\hbox{yr}}

\def\msun{M_{\odot}}

\def\ergcm2s{\ifmmode {\rm\,erg\,cm^{-2}\,s^{-1}}\else
                ${\rm\,ergs\,cm^{-2}\,s^{-1}}$\fi}
\def\kmsMpc{\ifmmode {\rm\,km\,s^{-1}\,Mpc^{-1}}\else
   ${\rm\,km\,s^{-1}\,Mpc^{-1}}$\fi}

\def\kms{{\rm km s^{-1}}}

\newcommand{\fluxunit}{\ergcm2s}

\begin{document}

\title{No X-ray Bright Type II Quasars Among the Lyman-$\alpha$ Emitters}

\author{S. Malhotra\altaffilmark{1}, J. X. Wang\altaffilmark{2}, J. E. Rhoads\altaffilmark{1}, T. M. Heckman\altaffilmark{2}, C. A. Norman\altaffilmark{1,2}}

\begin{abstract}

The \lya\ emitters found at z=4.5 and 5.7 by the Large Area Lyman
Alpha (LALA) survey have high equivalent widths in the \lya\
line. Such lines can be produced by narrow-lined active galactic
nuclei (AGNs) or by stellar populations with a very high proportion of
young, massive stars. To check for Type-II (i.e., narrow-lined)
quasars, we obtained a deep X-ray image of 49 \lya\ sources in a
single field of the ACIS instrument on the Chandra X-ray Observatory.
None of these sources were detected with a $3 \sigma$ limiting X-ray
luminosity of $2.9\times10^{43}$ ergs/s.  For comparison, the two
known high redshift type-II quasars have luminosities of $4 \times
10^{43}$ ergs/s before extinction correction.  The sources remain
undetected in stacked images of the 49 \lya\ sources (with 6.5 Ms
effective $Chandra$ on-axis exposure) at 3$\sigma$ limits of
$4.9\times10^{42}$.  The resulting X-ray to \lya\ ratio is about 4-24
times lower than the ratio for known type-II quasars, while the
average \lya\ luminosity of the LALA sample is in between the two
type-II's.  The cumulative X-ray to \lya\ ratio limit is also below that of
 90\% of low-redshift Seyfert galaxies.
\end{abstract}

\keywords{none supplied}

\altaffiltext{1}{Space Telescope Science Institute, 3700 San Martin Drive, Baltimore, MD 21218; san@stsci.edu, rhoads@stsci.edu} 
\altaffiltext{2}{Department of Physics and Astronomy, Johns Hopkins University, Baltimore, MD 21218; jxw@pha.jhu.edu, heckman@pha.jhu.edu, norman@stsci.edu}

\section {Introduction}

A large population of high redshift AGNs would have interesting
implications both for the pace of black hole formation and growth in
the universe and for cosmic background radiations from the gamma ray
to the far infrared.  Of particular interest is the possibility of a
large population of type II quasars, i.e., systems whose broad line
regions and soft x-rays are greatly attenuated by large column
densities of gas and dust.  Population synthesis models of active
galactic nuclei that are built to explain soft and hard X-ray source
counts and backgrounds predict that such objects comprise as much as
90\% of the high redshift quasar population (e.g., Gilli, Salvati, \&
Hasinger 2001).  The first X-ray selected type~II quasars
have recently been found (Norman et al 2002, Stern et al. 2002). These 
type~II quasars show prominent, narrow \lya\ emission lines, comparable in
luminosity (2-18$\times 10^{42}$ ergs/s) to the LALA sample ($>
4 \times 10^{42}$ ergs/s).

The equivalent widths of \lya\ emitters selected using narrow-band
surveys tend to be large (Malhotra \& Rhoads 2002 - hereafter MR02, Kudritzki et al. 2000). The median equivalent width of the \lya\ line is greater than
200 \AA| in a sample of 160 \lya\ emitters at z=4.5 and 18 at z=5.7 (MR02, RM01). Normal stellar populations can produce \lya\ emission with
equivalent width 240 \AA\ or less (Charlot \& Fall 1993), unless they
have a top-heavy initial mass function (IMF), zero (or very low)
metallicity, and/or extreme youth (age $< 10^7$ years). The high
equivalent widths could also be explained if active galactic nuclei
(AGNs) were present in our \lya\ emitter sample.  However, neither
narrow-band imaging nor spectroscopy shows evidence of broad emission
lines, which rules out classical quasars. Inspired by the recent
discovery of type-II quasars, we use deep x-ray imaging to search for
type~II quasars among the \lya\ emitters.

\section{Observations}

\section{Optical Data and Sample Selection}
\label{optobs}
The LALA survey comprises two fields, located in Bo\"{o}tes (at
14:25:57 +35:32 J2000.0) and in Cetus (at 02:05:20 -04:55 J2000.0).
Each field is $36 \times 36$ arcminutes in size, corresponding to a
single field of the 8192x8192 pixel Mosaic CCD cameras at the National
Optical Astronomy Observatory's 4 meter telescopes.  The X-ray
observations described in this {\it Letter\/} are in the Bo\"{o}tes
field.  In this field, we have LALA survey data in a total of eight
narrowband filters.  Five are partially overlapping narrow band
filters covering $4.37 < z < 4.57$ for \lya. There are also two
non-overlapping filters of similar width covering $5.67 < z < 5.80$
($\lambda_c \approx 8150$ and $8230$\AA).  Imaging data reduction
followed the methods described by Rhoads et al (2000), and \lya\
candidates were selected using criteria descibed by RM01 This resulted
in Bo\"{o}tes field samples of $\approx 160$ good candidates at
$z\approx 4.5$ (see Rhoads et al 2000; MR02) and $18$ at $z\approx
5.7$ (Rhoads \& Malhotra 2001). The Chandra field was placed to
maximize the number of large equivalent width sources within the
ACIS-I field of view.

\subsection{Chandra imaging}

A total of 178 kilo-second exposure, composed of two individual
observations, was obtained using the Advanced CCD Imaging Spectrometer
(ACIS) on the {\it Chandra X-ray Observatory\/} in very faint (VFAINT)
mode.  The first observation, with 120 ks exposure, was taken on
2002-04-16/17 ({\it Chandra\/} Obs ID 3130).  The second observation,
with 58 ks exposure, was taken on 2002-06-09 (Obs ID 3482).  All four
ACIS-I chips and ACIS-S2, ACIS-S3 chips were used,
with the telescope aimpoint centered on the ACIS-I3 chip
for each exposure. The
aimpoint of Obs ID 3130 is 14:25:37.791 +35:36:00.20 (J2000.0), and
the aimpoint of Obs ID 3482 is 14:25:37.564 +35:35:44.32,
16$\arcsec$ away from that of Obs ID 3130.
Due to their large off-axis angle during the observations, 
the ACIS-S chips have poorer spatial
resolution and effective area than the ACIS-I chips.
In this paper, data from any ACIS-S CCD were then ignored.

Data reduction was done with the package CIAO 2.2.1 (see
http://asc.harvard.edu/ciao).  The
level 1 data were reprocessed to clean the ACIS particle background
for very faint (VFAINT) mode observations, and filtered to include
only the standard event grades 0,2,3,4,6. All bad pixels and columns
were also removed.  We excluded high background time intervals
from level 2 files, leaving a net exposure time of 172 ks (120
ks from Obs ID 3130, and 52 ks from Obs ID 3482). Three images were
extracted from the combined event file: a soft image (0.5 -- 2.0 keV),
a hard image (2.0 -- 7.0 keV) and a total image (0.5 -- 7.0 keV). The
hard and total bands were cut at 7 keV since the effective area of
{\it Chandra\/} decreases above this energy, and the instrumental background
 rises, giving a very inefficient detection of sky and source
photons. 
The average offset between X-ray and optical images was obtained by
comparing the X-ray source positions and their optical counterparts 
(whenever found). Such offset (0.5$\arcsec$) has been corrected for
all our analysis in this paper. We ran WAVDETECT (Dobrzychi et
al. 1999, Freeman et al. 2002) on the soft, hard, and total band
images. A probability threshold of 1 $\times$ 10$^{-7}$ (corresponding
to 0.5 false sources expected per image), and scales of
1,2,4,8,16 pixels were used. 
The detection is down to a limiting flux of 1.6 $\times$ 10$^{-16}$
\ergcm2s in 0.5 -- 2.0 keV band, and 1.7 $\times$ 10$^{-15}$ \ergcm2s
in 2.0 -- 10.0 keV band.
Given the uncertainty on the value of the total background, 
we find that  $>$ 65\% of the hard X-ray background is resolved to a 
flux limit of 1.7 $\times$ 10$^{-15}$ \ergcm2s in 2.0 -- 10.0 keV band.
The detailed results and detected X-ray sources will be
published in a future paper (Wang et al. 2003).

\subsection{Non-detection of individual sources}

Forty nine of the Ly$\alpha$ sources were imaged by the $Chandra$
exposure with different effective exposure times.  For every X-ray
source and Ly$\alpha$ source, we defined a circular source region
centered at the source position and with radius R$_s$ set to the 95\%
encircled-energy radius of {\it Chandra\/} ACIS PSF at the position.
Note that at larger off-axis angle, we have larger PSF size.  Only one
Ly$\alpha$ source overlaps any of the X-ray source regions.  The 95\%
encircled-energy radius of $Chandra$ ACIS PSF at the overlapped
position is 10.2$\arcsec$, and the corresponding X-ray source and
Ly$\alpha$ source are 7.5$\arcsec$ apart; which is much larger than
the 3$\sigma$ positional error derived from X-ray image
(3.5$\arcsec$).  The number of X-ray pixels encircled by all the X-ray
detected source regions is around 80,000, so the probability that one
of the 49 Ly$\alpha$ sources fell in an X-ray source region is $\sim
8\times10^4/5\times 10^6 \times 49 \approx$ 78\%.  Thus, the possible
coincidence of one X-ray source with one \lya\ source is not
statistically significant.

We also performed X-ray photometry analysis of the 49 Ly$\alpha$
sources.  We again used the 95\% encircled-energy radius R$_s$
(now centered on the \lya\ coordinates) as the region to extract source
photons, and extracted the background
from an annulus with $1.2 R_s < R < 2.4 R_s$
after masking out nearby sources. We
also accounted for differences of exposure time between source regions
and background regions (mainly due to CCD edge effects and bad columns).
In soft band (0.5 -- 2.0 keV), the counts in the source regions for
all 49 sources are less than the 90\% significance level upper limits
of the expected background, with net counts all less than 2.6. In
total band (0.5 -- 7.0 keV), except for one source with net count of
7.8, all other sources have net counts $<$ 4, and confidence level $<$
90\%.  

The detection of the only source with 7.8 net counts in total band is
right at 2-$\sigma$ level.  Assuming a powerlaw spectrum with photon
index of 1.4 (from the average spectrum of all sources in the field),
this implies an X-ray flux of 6.3 $\times$ 10$^{-16}$ \fluxunit in 0.5
-- 10.0 keV. The source is detectable by WAVDETECT with a much lower
threshold level (10$^{-4}$), and the distance between the center of
the detected X-ray source and the Ly$\alpha$ source is 1.5$\arcsec$,
while the PSF value there is 9$\arcsec$.  However, with a threshold
level of 10$^{-4}$, we expect 0.15 false X-ray source (per X-ray image
from WAVDETECT) within 1.5$\arcsec$ of the 49 Ly$\alpha$ sources, so
the statistical significance of the detection remains low.  There is
no optical continuum counterpart for the possible X-ray source in our
broad band optical images. So if real, it should
be the counterpart of a Ly$\alpha$ source
with an rest-frame equivalent width of $>$500\AA.

We conclude that except one possible detection at 2$\sigma$ level 
in total band, none of the Ly$\alpha$ sources are detected by the
X-ray observations.  Only upper limits of X-ray fluxes of the
Ly$\alpha$ sources can be given.  For the Ly$\alpha$ source nearest
(1.8$\arcmin$) to the axis of X-ray observation, there are no photons
within R$_s$.  The 3$\sigma$ level upper limits of X-ray counts are
6.61 (Gehrels 1986), and the upper limits for X-ray fluxes (for
powerlaw spectra with photon index of 1.4) are 1.7 $\times$ 10$^{-16}$
\fluxunit for soft band, and 4.7 $\times$ 10$^{-16}$ \fluxunit for
total band (0.5 -- 10.0 keV). If we use photon index of 2, instead,
the above two fluxes will change to 1.9 $\times$ 10$^{-16}$ and 3.3
$\times$ 10$^{-16}$ respectively. Other sources have higher upper
limits due to the lower effective areas and larger PSF sizes.

\subsection{Cumulative X-rays from all the LALA sources}

The X-ray imaging data at the positions of all the \lya\ sources were
stacked, yielding an effective exposure time of 6.5 Ms (75 days). No
source was detected in the stacked image in any band (Figure 1).
Since Ly$\alpha$ sources have different off-axis angles to the axis of
{\it Chandra\/} observation, they have different PSF sizes at each
position, making it hard to define a source region to do photometry.
We just sum up the source counts extracted from each source region,
and the expected background counts derived from each background
region.  In the soft band, we have 55 counts in total in the source
regions, and the expected background is 54.2. In the total band, these
two numbers are 164 and 163.1 respectively. It is clear that we did
not detect the sources in X-ray even after stacking them. The
3$\sigma$ upper limits of soft and total band net counts are 26 and
42. Assuming a powerlaw spectrum with photon index of 1.4, we have
$3\sigma$ upper limits of 1.8 $\times$ 10$^{-17}$ \fluxunit in the 0.5
-- 2.0 keV band, and 7.9 $\times$ 10$^{-17}$ \fluxunit in the 0.5 --
10.0 keV band. For a photon index of 2, these limits translate to 1.9
$\times$ 10$^{-17}$ \fluxunit and 5.5 $\times$ 10$^{-17}$ \fluxunit.

\begin{figure}
\plotone{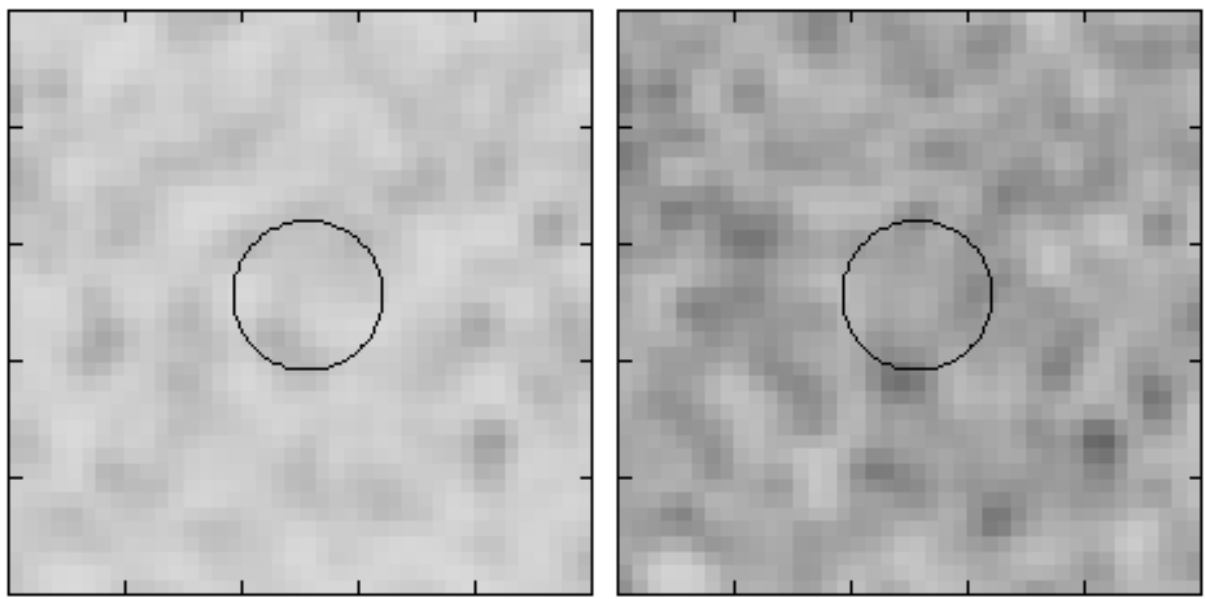}
\caption{Stacked $Chandra$ images of 49 Ly$\alpha$ sources. Left:
soft band (0.5 -- 2.0 keV); Right: total band (0.5 -- 
7.0 keV). The effective exposure time of the stacked images
is 6.5 Ms. The images are 40 $\times$ 40 pixels in size, and 
the circles are centered on the stacking position and have a radius of
5 pixels. 1 pixel = 0.492 $\arcsec$. The images were smoothed
using program provided by CIAO.}
\end{figure}

\section {Discussion and conclusions}

From the measurements of the X-ray background and from number counts
of X-ray sources, one can argue that only a small fraction of \lya\
emitters can be X-ray bright type-II quasars. Suppose a fraction $f$
of our sources were type-II quasars, i.e., much like CDF-S202 source
and distributed in redshift span $\Delta z$. The known type-IIs are at
z=3.3 and 3.7.  If these sources and the LALA objects at $z=4.5$ and
$5.7$ represent the same population, it would imply $\Delta z >
2.4$. The number counts of objects brighter than CDF-S202 and CXO52 in
the hard band in one ACIS field is about 160, while there are $49$
\lya\ sources at $z=4.47 \pm 0.1$ in the same solid angle.  This
implies immediately that not more than 27\% of the LALA sources should
be detectable in our X-ray survey. Even if we deem type~II quasars
irrelevant to this estimate, given the non-detection of LALA sources;
\lya\ sources have been found at z=3.1 with number densities
comparable to the LALA determinations ( Kudritzki et al. 2000) making
$\Delta z=2.6$.

Assuming we have a type II AGN at z=4.5, with intrinsic photon index
of 2, but heavily absorbed, $N_H = 10^{24}\hbox{cm}^{-2}$,
$H_0=65\kmsMpc$, $\Omega_m=1/3$, $\Omega_\lambda=2/3$.  The best
3-$\sigma$ upper limits of 0.5 -- 7.0 keV band net count at individual
LALA locations correspond s to a type II AGN with 2.0 -- 10.0 keV rest
frame intrinsic luminosity of 1.5 $\times 10^{44}$ ergs s$^{-1}$. And
for stacked images, this number is 2.5 $\times 10^{43}$ ergs
s$^{-1}$. The rest frame intrinsic luminosity of CDFS-202 is 8.0
$\times 10^{44}$ ergs s$^{-1}$ (Norman et al. 2002), and 4.2 $\times
10^{4 4}$ ergs s$^{-1}$ for CXO52 (Stern et al. 2001). See Table 1 for
corresponding luminosities with no extinction correction applied.

\begin{deluxetable}{lccr}
\tablecolumns{4}
\tablewidth{0pc} 
\tablecaption{X-ray flux and Luminosity measurements and limits}
\tablehead{
\colhead{Object Type} & \colhead{Flux} & \colhead{Luminosity (2-10 keV)}\\
\colhead{} & \colhead{\fluxunit} &\colhead{ ergs s$^{-1}$}}
\startdata 
Individual \lya\ emitters &  $<3.3 \times 10^{-16}$ &  $< 2.9\times 10^{43}$\\
Coadded \lya\ emitters &  $<5.5 \times 10^{-17}$& $<4.9 \times 10^{42}$\\
Type II quasar: CDF-S202 & $2.3-2.8 \times 10^{-15}$ & 4.4 $\times 10^{43}$\\
Type II quasar: CXO-52 & 1.8 $\times 10^{-15}$& 3.6 $\times 10^{43}$\\
Lyman Break Galaxies (Coadded non-AGN) & 4 $\times 10^{-18} $& $6 \times 10^{40}$ \\
\lya\ galaxies if starbursts & & $6\times 10^{39}$  \\
\enddata 
\end{deluxetable} 

However, if we were to scale the X-ray fluxes with \lya\ line flux,
which in most cases is the only well measured property of the LALA
sources, we find that about 44 out of 49 sources would have been
detected had they been like CDF-S202. About 3 sources like CXO52 would
have been detected by these observations. Figure 2a shows the
comparison, where the ratio of X-ray 3$\sigma$ upper limits to \lya\
line flux for LALA sources is shown as a histogram and the values of
this ratio for the two known type-II quasars are marked. The spread in
the histogram is due to variation in x-ray flux sensitivity and the
variation in \lya\ line strength. Also shown is this ratio for
cumulative LALA source positions which is 4-24 times lower than either
of the known QSO-IIs.  The 3-$\sigma$ upper limit of average X-ray
flux from each source is $5.5 \times 10^{-17}$ \fluxunit in the
0.5--10 KeV band, compared to a few $\times 10^{-15}$ \fluxunit for
the two known type-II quasars.  Comparison with relatively lower
luminosity, low redshift ($z<1$)Seyferts and quasars drawn from the
samples of Kriss (1984, 1985) shows that while individually the LALA
upper limits are higher than the X-ray/\lya\ ratio for about half the
low-z sample, the upper limits from the stacked LALA positions is
lower than 90\% of low-z sources (Figure 2b).


Assuming a power-law spectrum with photon index $\Gamma$ = 2.0, the
$3\sigma$ upper limit of 0.5 -- 2.0 keV flux on the average LALA
sources corresponds to an X-ray luminosity of $4.2 \times 10^{42}$
ergs s$^{-1}$ at z=4.5 (for either 0.5 -- 2.0 keV rest frame or 2.0 --
8.0 keV rest frame bandpass, H$_0$=65 km s$^{-1}$ Mpc$^{-1}$,
$\Omega_m$=1/3, $\Omega_\lambda$=2/3).  The Lyman break galaxies at $z
\simeq 3$ have an average luminosity of $6 \times 10^{40}\erg s^{-1}$
(after excluding the four known AGNs) which is consistent with their
being starbursts with star-formation rates of about 60 $\msun
\year^{-1}$ (Nandra et al 2002, Brandt et al. 2001). Clearly, our
observations are not sensitive enough to detect starbursts (nor were
they designed to be).  Taking the difference in star-formation rates
between \lya\ selected galaxies and LBGs into account and adopting the
LBG value for the ratio of star formation to X-ray emission, we should
expect an average x-ray luminosity of $\approx 6\times 10^{39} \erg
s^{-1}$.

Could some of the Lyman-$\alpha$ sources still be AGNs?  We
have demonstrated that X-ray bright quasars are at most a minority of
the LALA objects.  Our composite nondetection implies that even at the
3$\sigma$ level, only a few percent of LALA sources could resemble
CDF-S202, and $<25\%$ could resemble CXO52.  The most plausible
way to have luminous AGN hiding in the LALA sample without violating this
constraint is to suppose that they are Compton-thick, so that even
relatively hard X-rays are obscured.  Thermal emission from the obscuring 
dust would render these objects detectable in the infrared or submillimeter.

Another test relies on optical spectra.  The \lya\ line is found to be
narrow ($< 500 \kms$) in all our spectroscopically confirmed \lya\
emitters (e.g., Rhoads et al. 2000, Rhoads et al. 2002), which is
narrower than the typical physical line widths of even type II
quasars.  For the larger fraction of \lya\ sources which are
photometrically selected using narrow-bands 80 \AA\ wide, we can also
rule out velocities $ >3700 \kms$.  Steidel et al (2002) find at least
one case of narrow-lined AGN in their Lyman-break galaxy sample with
x-ray luminosity $< 5 \times 10^{42}$ ergs~s$^{-1}$. Their
identification of this source as an AGN is based on the detection of
narrow lines of NV, CIV, HeII and CIII in emission.  None of the
spectra of \lya\ emitters shows these lines. In conclusion, we find no
evidence for AGN among the \lya\ emitters found in the LALA survey.

\acknowledgements 
This work has benefitted from images provided by the NOAO Deep Wide-Field
Survey (NDWFS; Jannuzi and Dey 1999), which is supported by the National
Optical Astronomy Observatory (NOAO).  NOAO is operated by AURA,
Inc., under a cooperative agreement with the National Science
Foundation. We also thank the referee for a prompt and helpful report.

\clearpage

\begin{figure}
\plottwo{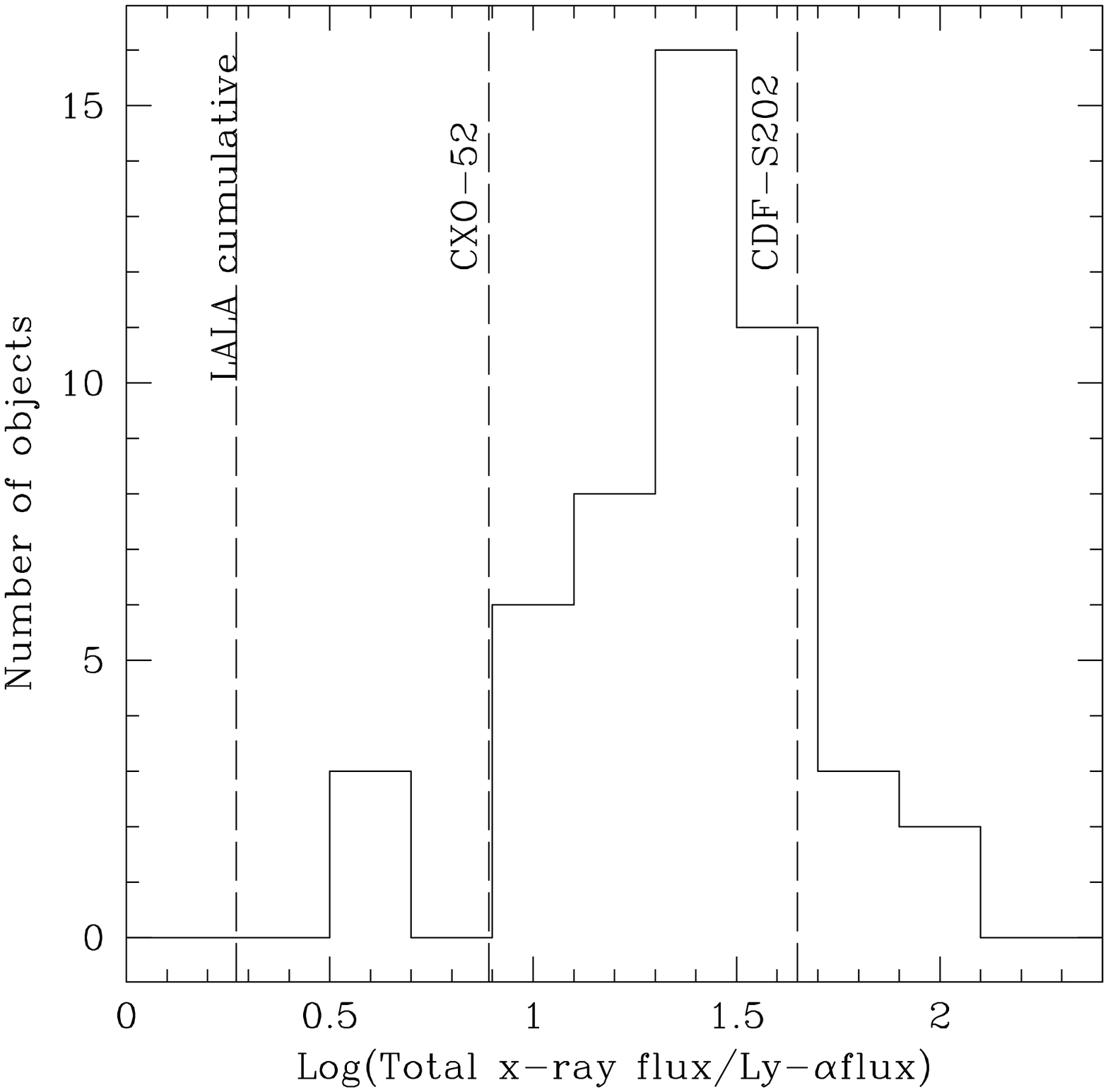}{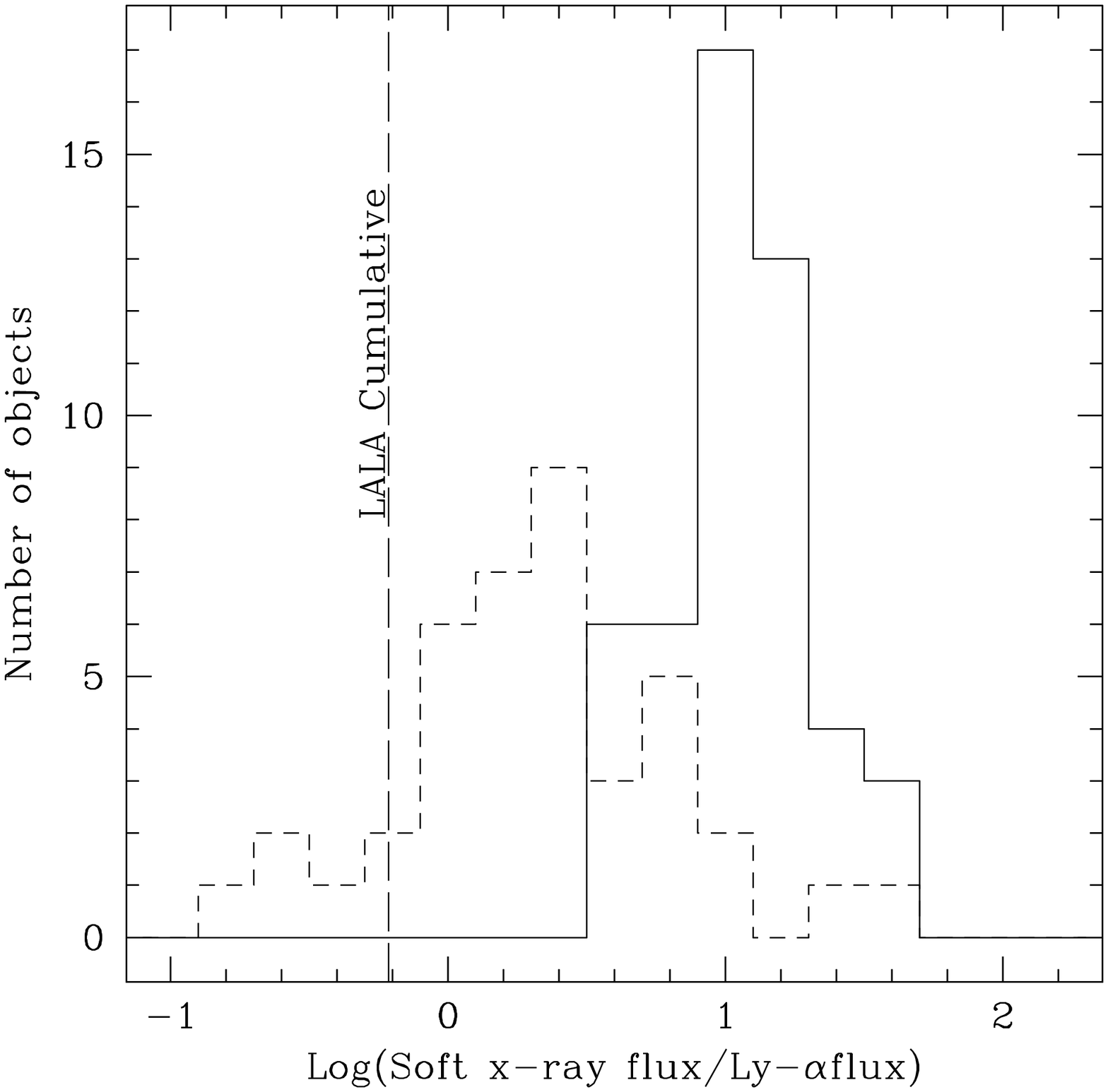}
\caption{ Comparison of X-ray to Lyman-$\alpha$ flux ratios: Figure 2a
shows the comparison of this ratio for the 49 \lya\ emitters at z=4.5
from the LALA sample with high redshift Type-II quasars. We use 3-$\sigma$ upper limits derived for the
total band since there are no detections. The vertical bars mark the
observed ratio for the two known high redshift type-II quasars (see
text), and the 3-$\sigma$ upper limit for all the \lya\ sources
stacked. Since the photon indices of the known type-II quasars are
different from the assumed photon index for the undetected \lya\
emitters, we have adjusted the x-ray fluxes in all sources as if the
photon index were $\Gamma=2$.  Figure 2b shows comparison with
low-redshift ($z<1$) Seyfert galaxies and quasars (Kriss 1984, 1985).
The solid histogram is as in Figure 2a, except we plot the soft band
(0.5--2.0 keV, observed; 2.75--11 keV, rest) and the dashed histogram is the comparison
sample from Kriss (1984 \& 1985) adjusted to rest-frame 2.75--11
keV. The 3-$\sigma$ upper limit on the stacked \lya\ sources is shown
as a vertical bar.}
\end{figure}

\end{document}